\documentclass{article} 
\usepackage{nips13submit_e,times}
\usepackage{url}
\usepackage{graphicx,float}
\usepackage{amssymb,amsmath,amsthm}
\usepackage{rotating}
\newcommand{\ignore}[1]{}
\title{Entropy Dynamics of Community Alignment in the Italian Parliament Time-Dependent Network}
\author{
Gabriele Lami\\
PolicyBrain, Milano (Italy)\\
\texttt{gabriele.lami@policybrain.com} \\
\And
Marco Cristoforetti\\
ECT*, Trento (Italy)\\
\texttt{cristoforetti@gmail.com} \\
\And
Giuseppe Jurman\\
FBK, Trento (Italy)\\
\texttt{jurman@fbk.eu} \\
\And
Cesare Furlanello\\
FBK, Trento (Italy)\\
\texttt{furlan@fbk.eu} 
\And
Tommaso Furlanello\thanks{Corresponding author}\\
USC, Los Angeles, CA (US)\\
\texttt{furlanel@usc.edu} \\
}

\newcommand{\bm}[1]{\bf{#1}}
\nipsfinalcopy 
\begin{document}
\maketitle
\begin{abstract}
Complex institutions are typically characterized by meso-scale structures which are fundamental for the successful coordination of multiple agents.
Here we introduce a framework to study the temporal dynamics of the node-community relationship based on the concept of \textit{community aligment}, a measure derived from the modularity matrix that defines the alignment of a node with respect to the core of its community. 
The framework is applied to the 16th legislature of the Italian Parliament to study the dynamic relationship in voting behavior between Members of the Parliament (MPs) and their political parties. 
As a novel contribution, we introduce two entropy-based measures that capture politically interesting dynamics: the \textit{group alignment entropy} (over a single snapshot), and the \textit{node alignment entropy} (over multiple snapshots).
We show that significant meso-scale changes in the time-dependent network structures can be detected by a combination of the two measures. We observe a steady growth of the group alignment entropy after a major internal conflict in the ruling majority and a different distribution of nodes alignment entropy after the government transition.
\end{abstract}

\section{Introduction}
\label{sec:intro}
Large voting assemblies produce huge amount of data on the behavior of their constituents and on their organization into groups and coalitions. 
Previous studies modeling assemblies as complex network focused on their community structure revealing the influence of the committees organization on roll call voting in the U.S. House of Representatives ~\cite{porter05network}, and how national and geopolitical events are associated with long term variations of the community structure of the United States Congress~\cite{waugh11party} and the United Nations General Assembly~\cite{macon12community}. 
Previous analysis on the Italian Parliament \cite{amelio12analyzing,dalmaso2014voting} confirmed the association between the party structure and the community structure as revealed by algorithms based on modularity maximization~\cite{newman06modularity}.
In this article we focus on the distinctive characteristic of the Italian assembly of being composed by instable coalitions that rarely survive the mandate to study the dynamic relationship among nodes and their communities. 
We model the Italian Parliament as a series of temporal networks where nodes correspond to MPs and links are defined by their similarity in voting behavior over a time window of 30 days. 
As in precedent studies, we consider a spectral reformulation of the modularity matrix \cite{newman06finding,waugh11party} as our source of information on the community structure, allowing us to to define the community alignment of each node, in our case corresponding to the alignment of each MP to its political party over time. 
We use the entropy in community alignment of each group to define a measure of partisan cohesion and the entropy of MPs alignment over time to define a measure of individual stability. 
In order to test the validity of our approach we aim at matching the quantitative results emerging from the analysis with the political events landmarking the period.\\
Section \ref{sec:methods} introduces our framework and presents a definition of community alignment and of the two entropy-based measures. 
Section \ref{sec:background} provides a brief political background on the 16th${}^\textrm{th}$ legislature of the Italian Parliament and Section \ref{sec:results} presents the results of our analysis during that period. 
Section \ref{sec:conclusions} concludes the article.
\section{Political background}
\label{sec:background}
In what follows, the strategy is applied to the Chamber of Deputies of the Italian Parliament during the 16th legislature.
The considered period coincides with the fourth government led by Silvio Berlusconi, who was in charge as premier from May, the 8th 2008 to November, the 16th 2011 and the government led by Mario Monti following the one of Berlusconi and lasted until April, the 28th 2013.
The first of the two governments was supported by a center-right coalition where Berlusconi's party, the Popolo della Libert\`a (PDL for short) was by far the largest component.
The PDL itself was generated as a fusion of two parties: the liberal-conservative Forza Italia, with Silvio Berlusconi as the leader and the right-wing Alleanza Nazionale (AN for short), led by Gianfranco Fini, President of the Chamber of Deputies during the 16th legislature.
From 2009 onwards, Fini moved progressively away from Berlusconi and the government's positions, with harsh political clashes occurring more and more frequently.
On 29 July 2010 Fini was accused by PDL's assembly of being unable to neutrally perform his duties as President of the Chamber of Deputies and was asked to resign.
The day after, Fini announced in a press conference the formation of a separate group from the PDL both in the Chamber and the Senate under the name Futuro e Libert\`a (FLI for short), confirming its support to Berlusconi's government. 
During the following 16 months, the FLI senators and deputies behaved more and more independently from their PDL colleagues; this was a key factor contributing to the fall of the Berlusconi government (November, the 16th 2011), with Mario Monti indicated as the new Italian premier. 
FLI with the UDC of Pier Ferdinando Casini and the API led by Francesco Rutelli joined together in the Nuovo Polo per l'Italia, the so called Terzo Polo (third pole) on December the 15th of 2010 showing a willingness of independence both from the right and left wing of the parliament.
\section{Methods}
\label{sec:methods}
\paragraph{Network Construction}
Operatively, the voting data of the 630 deputies (as retrieved from the official repository at the Italian Chamber of Deputies website \url{dati.camera.it}) are organized in a time series of 83 networks. Each of them is the result of the analysis of the votes in a period of one month and using a sliding window of 10 days between two consecutive intervals. 
The period considered is from May 2008 until December 2012.
Each network has the deputies as nodes $v_i$, while the edges $e_{ij}$ are defined by the following rule:
\begin{itemize}
\item Consider a deputy $v_i$ and a vote $V_k$, and rate the corresponding outcome as follows:
\begin{displaymath}
v_i(V_k) =
\begin{cases}
1 & \textrm{Yes}\\
-1 & \textrm{No}\\
0 & \textrm{Absent}\\
0.3 & \textrm{Abstained};
\end{cases}
\end{displaymath}
Indeed the abstention is not a neutral political act: the value for the abstention used in this paper is chosen in order to maximize the number of politically meaningful communities as emerging from the modularity analysis.
Most of the times, the political reason for absenteeism is not related to the vote itself, yielding that the best value for the absenteeism is zero.
\item Let $(V_1,\ldots,V_n)$ be the vector of all voting that took place in a given time interval;
\item Let $v_i(V_1^n)=(v_i(V_1),\ldots,v_i(V_n))\in \{1,-1,0,0.3\}^n$ the vector of all the choices of deputy $v_i$ on all the voting $V_k$ held in the given time interval;
\item Let $e_{ij}= 1- \frac{|v_i(V_1^n)-v_j(V_1^n)|}{2n}$ the weighted link $e_{ij}$ connecting deputies $v_i$ and $v_j$ in the given time interval.
\end{itemize}
Our approach relies on combining the classical theory of modularity and community detection in complex networks~\cite{newman06modularity,newman06finding}, with the concept of entropy, here measured over the community vectors.
The modularity function $Q$ for community detection is defined by the relation 
\begin{equation}
\label{eq:mod_gen}
Q=\# \{ \textrm{true edges within communities}\}  - \# \{\textrm{edges expected in a null model distribution} \}\ .
\end{equation}
Suppose now that the chosen community detection algorithm~\cite{lancichinetti09community,lancichinetti14erratum} detects $c$ communities $C_1,\ldots,C_c$, and let $G_k$ be the set of nodes belonging to the $k$-th community $C_k$.
Given the adjacency matrix $A_{ij}$ and a null model with probability $P_{ij}$ for the edge $e_{ij}$ between every pair of vertices $v_i,v_j$, the definition of modularity in Eq.~\ref{eq:mod_gen} reads as:
\begin{equation}
\label{eq:mod_1}
Q=\frac{1}{2m}\sum_{ij}\left[A_{ij}-P_{ij}\right]\delta(g_i,g_j)\ ,
\end{equation}
where $g_i$ is the community to which vertex $v_i$ belongs, $\displaystyle{m=\frac{1}{2}\sum_{ij}A_{ij}}$ is the number of edges in the network and $\delta$ is the Kronecker function $\delta(r,s)=\begin{cases} 1 & \textrm{if $r=s$} \\ 0 &\textrm{otherwise}\end{cases}$.
Let now $\partial(v_i)=\sum_j A_{ij}$ be the degree of vertex $v_i$: then, following Newman in~\cite{newman06modularity}, the expected number of edges $e_{ij}$ between vertices $v_i$ and $v_j$ if edges are placed at random is 
\begin{displaymath}    
P_{ij}=\frac{\partial(v_i)\partial(v_j)}{2m}\ .
\end{displaymath}
Clearly, more complex null models will give rise to different probability matrices $P_{ij}$~\cite{sarznyska14null}.
Finally, define the modularity matrix $B$ as $B_{ij}=A_{ij}-P_{ij}$, a real symmetric matrix that can be diagonalised with both real (positive, negative and zero) eigenvalues and eigenvectors. 
\paragraph{Community vectors}
In particular, let $n$ be the dimension of the matrix $B$, with eigenvalues $\beta_1 \geq \ldots \geq \beta_n$ and eigenvector matrix $\mathbf{U}=(u_1|...|u_n)$. 
Suppose then 
\begin{equation}
\beta_i = \begin{cases} >0 & \textrm{ for }1\leq i \leq p \\ 0 &  \textrm{ for }p+1\leq i\leq n-q \\ <0 &  \textrm{ for }n-q+1 \leq i\leq n \end{cases}
\end{equation}
with $p+q<n$. Following~\cite{newman06finding}, define two sets of vertex vectors $\{\bm{x}_i\}$ and $\{\bm{y}_i\}$ of dimension $p$ and $q$ respectively:
\begin{displaymath} 
\begin{split}   
\left\{\bm{x}_i\right\}_j&=\sqrt{\beta_j}\cdot U_{ij},\\
\left\{\bm{y}_i\right\}_j&=\sqrt{-\beta_{n+1-j}}\cdot U_{i,n+1-j}\ .   
\end{split}
\end{displaymath}
In terms of the vectors $\bm{x}$ and $\bm{y}$ the modularity in Eq.~\ref{eq:mod_1} can be rewritten as
\begin{equation}
\label{eq:cvmod}
Q=\sum_{k=1}^c\left(\left|\sum_{i\in G_k}\bm{x}_i\right|^2-\left|\sum_{i\in G_k}\bm{y}_i\right|^2\right)\ .
\end{equation}
In this form, Eq.~\ref{eq:cvmod} highlights the positive and negative contributions to the modularity.
Define now the community vectors $\bm{X}_k$ and $\bm{Y}_k$ as
\begin{displaymath}
\bm{X}_k=\sum_{i\in G_k}\bm{x}_i,\; \bm{Y}_k=\sum_{i\in G_k}\bm{y}_i\ .
\end{displaymath}
Decompose now $\left|\bm{X}_k\right|$ as
\begin{displaymath}
\left|\bm{X}_k\right| = \frac{\bm{X}_k^T\bm{X}_k}{\left|\bm{X}_k\right|}= \frac{\bm{X}_k^T}{\left|\bm{X}_k\right|} \sum_{i\in G_k}\bm{x}_i = \sum_{i\in G_k} \hat{\bm{X}}_k^T\bm{x}_i\ ,
\end{displaymath}
where $\hat{\bm{X}}_k$ is the unit vector in the direction of $\bm{X}_k$, so that each vertex vector gives a contribution to $\left|\bm{X}_k\right|$ equal to its projection onto $\bm{X}_k$. \\
\begin{figure}[!t]
\begin{center}
\begin{tabular}{cc}
\includegraphics[width=0.5\textwidth]{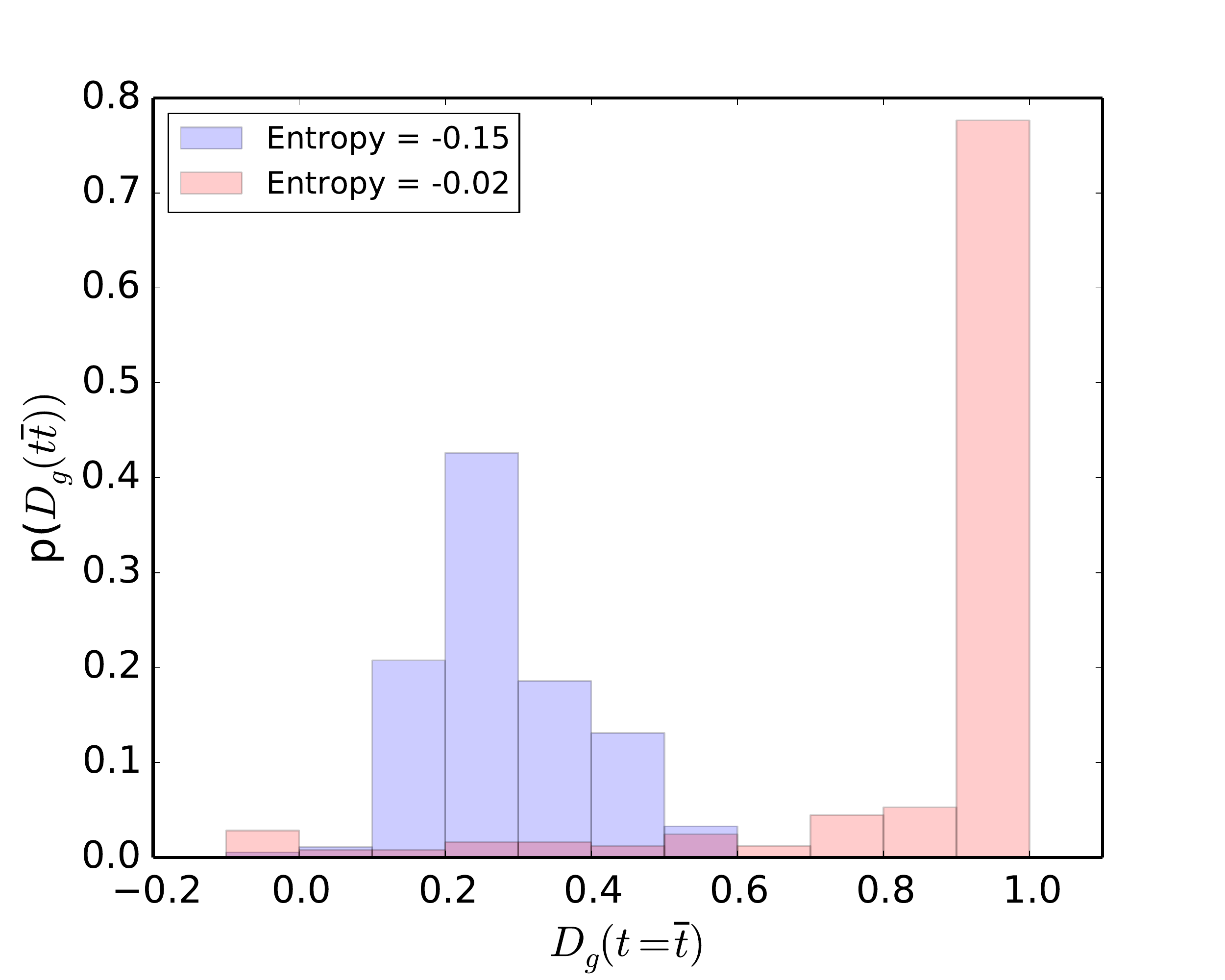} &
\includegraphics[width=0.5\textwidth]{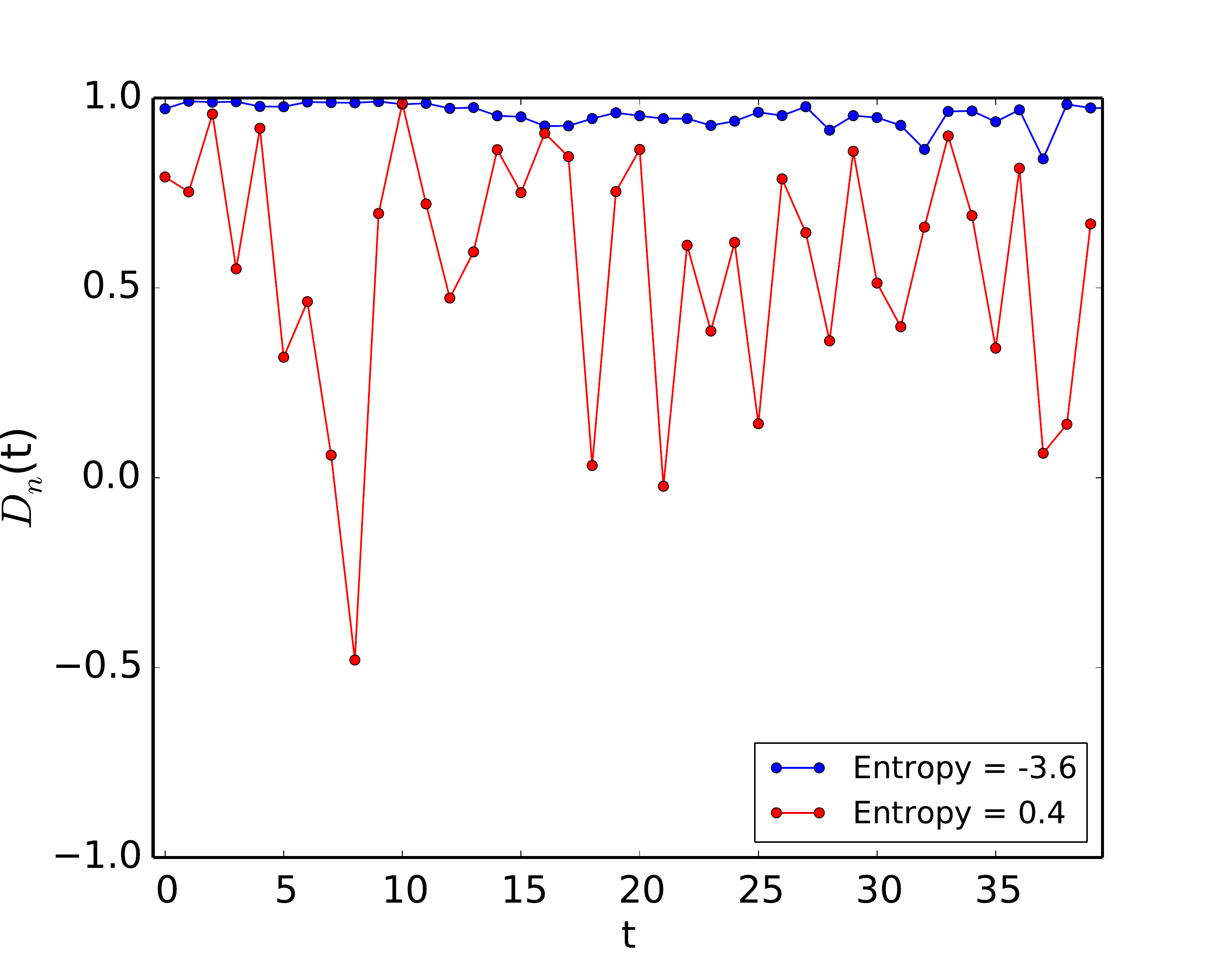} \\
(a) & (b)
\end{tabular}
\end{center}
\caption{(a) Group entropy: distribution of Community Alignment, at a timestamp $t=\overline{t}$, for two groups of MPs that return extreme values for the Group Alignment Entropy. (b) Node entropy: Community Alignment close to one on the whole timeframe return a large negative value for the Node Alignment Entropy (blue points), while the oscillating behaviour (red points) is instead an indicator of disorder and thus the associated entropy is close to zero.}
\label{fig:gn_entropy}
\end{figure}
\paragraph{Community Alignment}
A natural definition of community alignment derives from the angle $\theta_{ik}$ between the individual vector $\bm{x}_i$ and the community vector $\bm{X}_k$; we consider $\cos\theta_{ik}$ as a measure of the vertex's position with respect to the community.
In particular a community alignment \mbox{$\cos\theta_{ik}\approx 1$} occurs when $\bm{x}_i$ is in the core of the community, while a positive value close to zero indicates that $\bm{x}_i$ is at the periphery of community $\bm{X}_k$; negative values mark the case when $\bm{x}_i$ does not belong to $\bm{X}_k$.
In our study we focus on calculating $\cos\theta_{ik}$ for MPs in relation with their declared group of membership.
\begin{figure}[!b]
\begin{center}
\includegraphics[width=\textwidth]{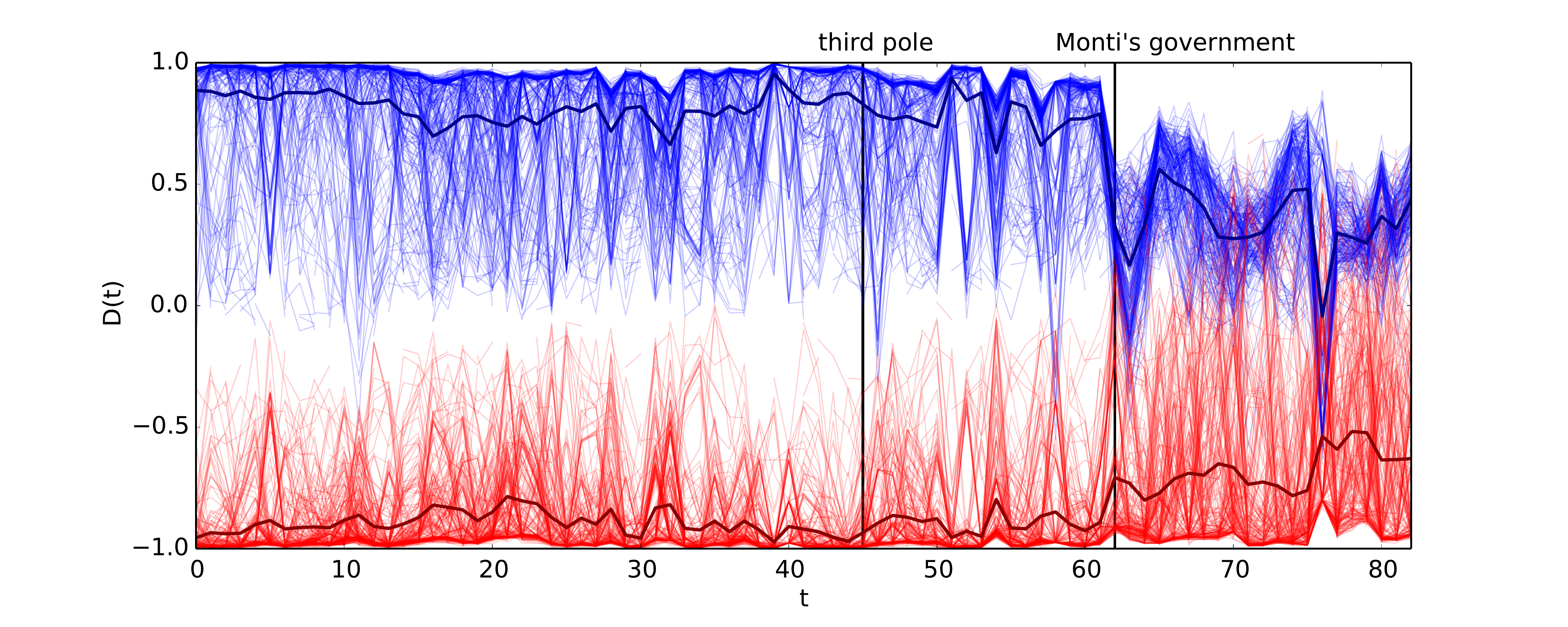}
\end{center}
\caption{Trajectories $D(t)$ for the deputies beloging to the PDL (blue) and PD (red); deputies later joining FLI are not considered after point 44. $D(t)$ near 1 indicates high alignment to the ruling coalition, conversely $D(t)$ near -1 indicates alignment to the opposition coalition}
\label{fig:costheta_reduced}
\end{figure}
\paragraph{Entropy}
Introducing the concept of entropy measure in the analysis of the distribution of the community vectors is useful in two ways: for a fixed timestamp, the community alignment entropy will provide an estimate of the disorder inside the groups. Symmetrically, at the level of the single deputies, given a set of consecutive timeframes we can compute the entropy for that period. In this case, we seek to identify a tendency of the MP to vote independently from the decision of the majority of her/his party. Given the general form of the entropy function $H$
\begin{displaymath}
H(X)=-\int \mu(x)\log\mu(x)\textrm{d}x\ ,
\end{displaymath}
for a random variable $X$ with distribution $\mu$, we use the approach in~\cite{kraskov04estimating} to derive its estimation for a random sample $(x_1,\ldots,x_N)$ of $N$ realizations of $X$: 
\begin{equation}
\label{eq:en}
\hat{H}(X)=-\psi(k)+\psi(N)+\log c_d+\frac{d}{N}\sum_{i=1}^N\log \epsilon(i),
\end{equation}
where $\psi=\Gamma(x)^{-1}\textrm{d}\Gamma(x)/\textrm{d}x$, $k$ is the order of the considered neighbor, $d$ is the dimension of $x$, $c_d$ is the volume of the $d$-dimensional unit ball (whose value depends on the adopted metric) and $\epsilon (i)$ is twice the distance from $x_i$ to its $k$-th neighbor.

\begin{figure}[!b]
\begin{center}
\includegraphics[width=\textwidth]{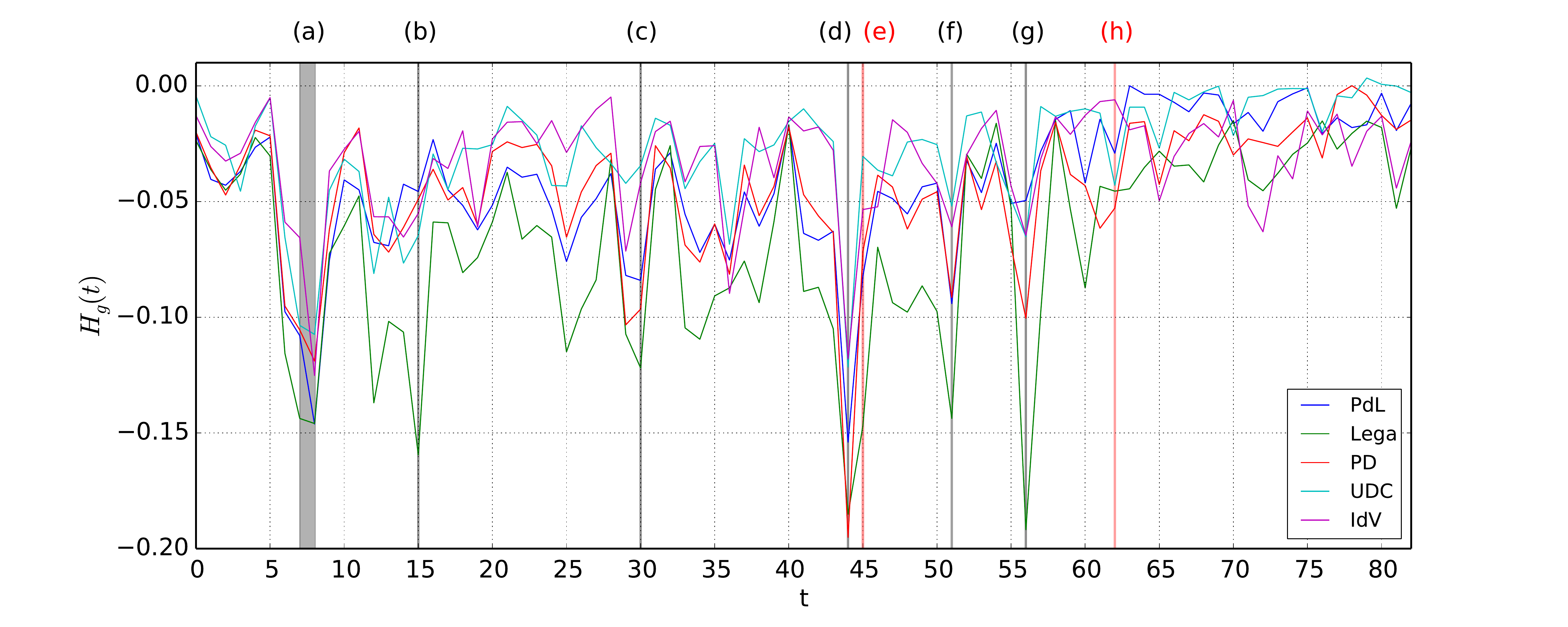}
\end{center}
\caption{Entropy for the five larger parties inside the Chamber of Deputy during the XVI legislature, normalized with respect to the number of votes in the given timeframe. Significant political events corresponding to spikes in the entropy are marked by a vertical grey line: (a) \textit{riforma Gelmini} - school reform, \textit{decreto Alitalia} - Alitalia recovering manouvres, \textit{salva banche} - bank bailout. (b) \textit{federalismo fiscale} - fiscal federalism. (c) International peace mission funding, \textit{milleproroghe} - omnibus law funding a number of spending extensions. (d) Stability Law 2011. (f) Short trial, Nuclear plants. (g) Elimination of \textit{province} local entities, corrections to stability law, peace mission fundings. Two additional events correspond to dramatical changes in the composition of the majority and opposition: (e) Birth of the Third Pole (h) Establishment of Monti's government.}
\label{fig:e1}
\end{figure}
\paragraph{Entropy alignment measures}
Applying Eq.~\ref{eq:en} on a single timepoint $t=\overline{t}$ we can compute the \textit{group alignment entropy} by studying the distribution of $\cos\theta$ for a given community of $N$ deputies $X^{\overline{t}}= [x_1^{\overline{t}},...,x_N^{\overline{t}}]$, while, considering only a single deputy, the \textit{node alignment entropy} can be computed by looking at $\cos\theta$ associated with the community vectors $X_i= [ x_i^{t_1},..,x_i^{t_N}]$ for the sequence of timestamps $t\in[t_1,...,t_N]$.
Examples of the group and node entropy measures from data of 16th legislature are presented in Fig.~\ref{fig:gn_entropy}.
\section{Results}
\label{sec:results}
Community vector projections and entropy were computed on the Chamber of Deputies' voting activities, both at political group level and at single deputy level.
In what follows we show the most relevant findings, focussing on the five major parties, which at the beginning of the legislature were organised as the majority formed by the PdL (including the deputies that will later found the FLI group) and the Lega, while the opposition is represented here by PD, UDC and IdV.

\begin{figure}[!t]
\begin{center}
\includegraphics[width=\textwidth]{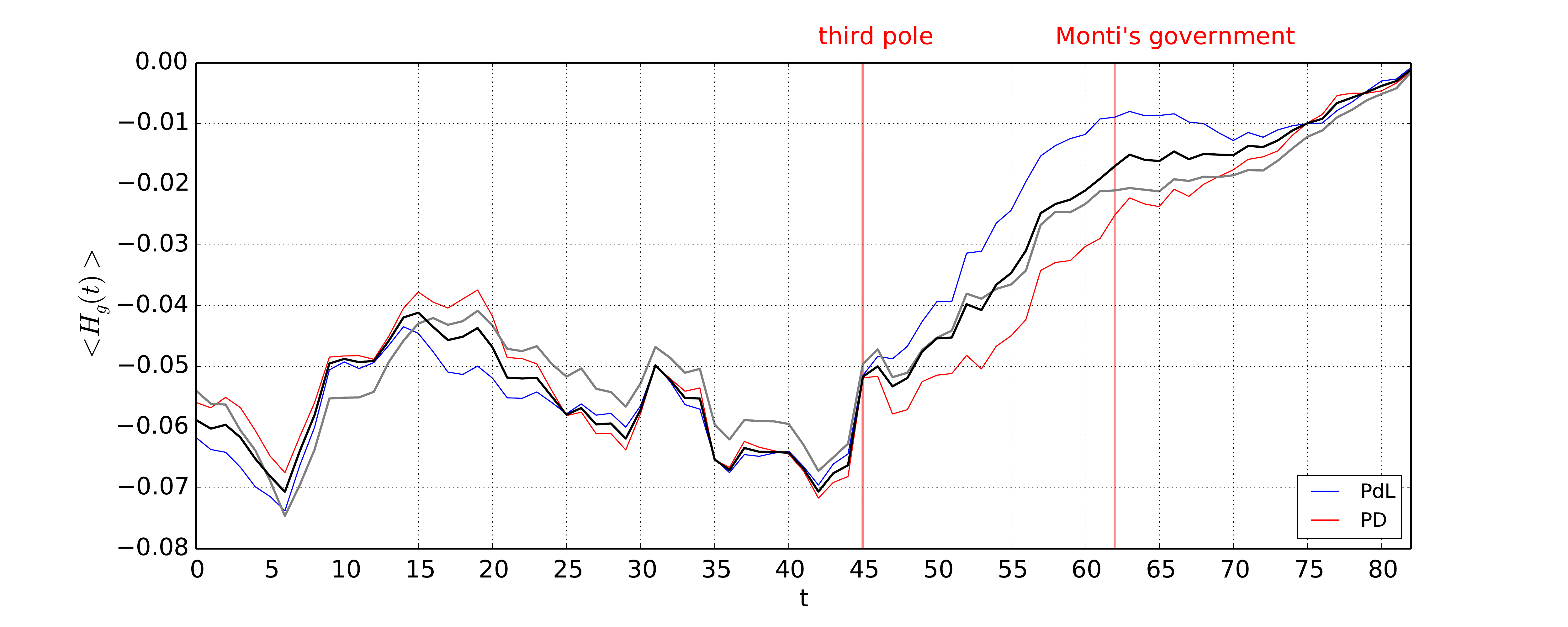}
\end{center}
\caption{Group entropy average on a sliding window with 10 timepoints: in blue PDL, in red PD, in black the PD/PDL average and in gray the average for all the Chamber of Deputies.}
\label{fig:e2}
\end{figure}

\paragraph{Projection on community vector}
Consider the deputies labelled as belonging to the ruling majority or to the opposition, according to the official structure for point 1.
This labeling is sufficiently accurate because immediately after the appointment of the Deputies the antagonism between the two competing coalitions was still harsh and no changes of team happened yet.
The curve $D_i(t)$, for $1\leq t\leq 83$ defines the trajectory of each deputy between the government and the opposition communities during the considered period. 
A deputy $v_i$ always voting following her/his party's guidelines has trajectory $D_i(t)=1$ (if $v_i$ is labeled M) or $D_i(t)=-1$ (if $v_i$ is labeled O). 
In Fig.~\ref{fig:costheta_reduced} we collect the trajectories for all deputies from PdL (excluding FLI) and PD.
It is interesting to notice that in the Monti's government (timepoints 62 to 82) the majority and opposition were totally different from the precedent period and this is reflected by the strong oscillation in the plot showing that the assignment by original coalitions do not coincide with the effective modular structure for these political networks.
\paragraph{Entropy analysis}
As stated at the beginning of the section, in our analysis we consider the five main groups present in the Chamber of Deputies: PD, PdL, Lega, UDC and IdV. 
As a first result, we compute the group entropy for the five larger parties inside the Chamber of Deputy during the XVI legislature, normalized with respect to the number of votes in the given timeframe. 
It is interesting to note how spikes in the group entropy correspond to significant political events, such as the stability law, a critical step for the national economy and usually a stress test for the cohesion of the majority.
Being the spike themselves associated to large negative value of the group entropy, this implies that, following the aforementioned events, the internal cohesion in the parties typically increases in particular for specific topics. For instance, the Lega's entropy keeps high negative values during the discussion of the fiscal federalism (the green line at the point (b) in the plot), a key point for the political platform for the party.
Clearly, a single negative spike can also identify a vote of confidence requested by the government.
Moreover, note that, from point 62 onwards (\textit{i.e.}, the establishment of Monti's goverment), for all parties the group entropy shows no more spikes and all values are close to zero, highlighting a period where many MPs were voting independently from their voting instructions.
To further investigate along this direction, we analyze in Fig.~\ref{fig:e2} the dynamics of the group entropy for the two major parties PdL and PD, by computing its average over a sliding window of 10 timepoints, with a mutual overlap of 9 points.
The trend of the entropy curve for the PD splits the time range into two well marked zones, with the former ending at point 44, and the latter corresponding to the final weeks of the legislation after the birth of the Third Pole.
In particular, while the first phase is characterized by an oscillating behavior of the entropy, always lower than -0.35, after point 45 entropy is nearly linearly increasing from -0.05 to 0.
For the PDL, the zones are instead three, since after point 62, its entropy tends to stabilize with the Monti's government (up to the final three months of the legislation): such trend is also shared by the average entropy of the whole Chamber.

\begin{figure}[!b]
\begin{center}
\begin{tabular}{cc}
\includegraphics[width=0.5\textwidth]{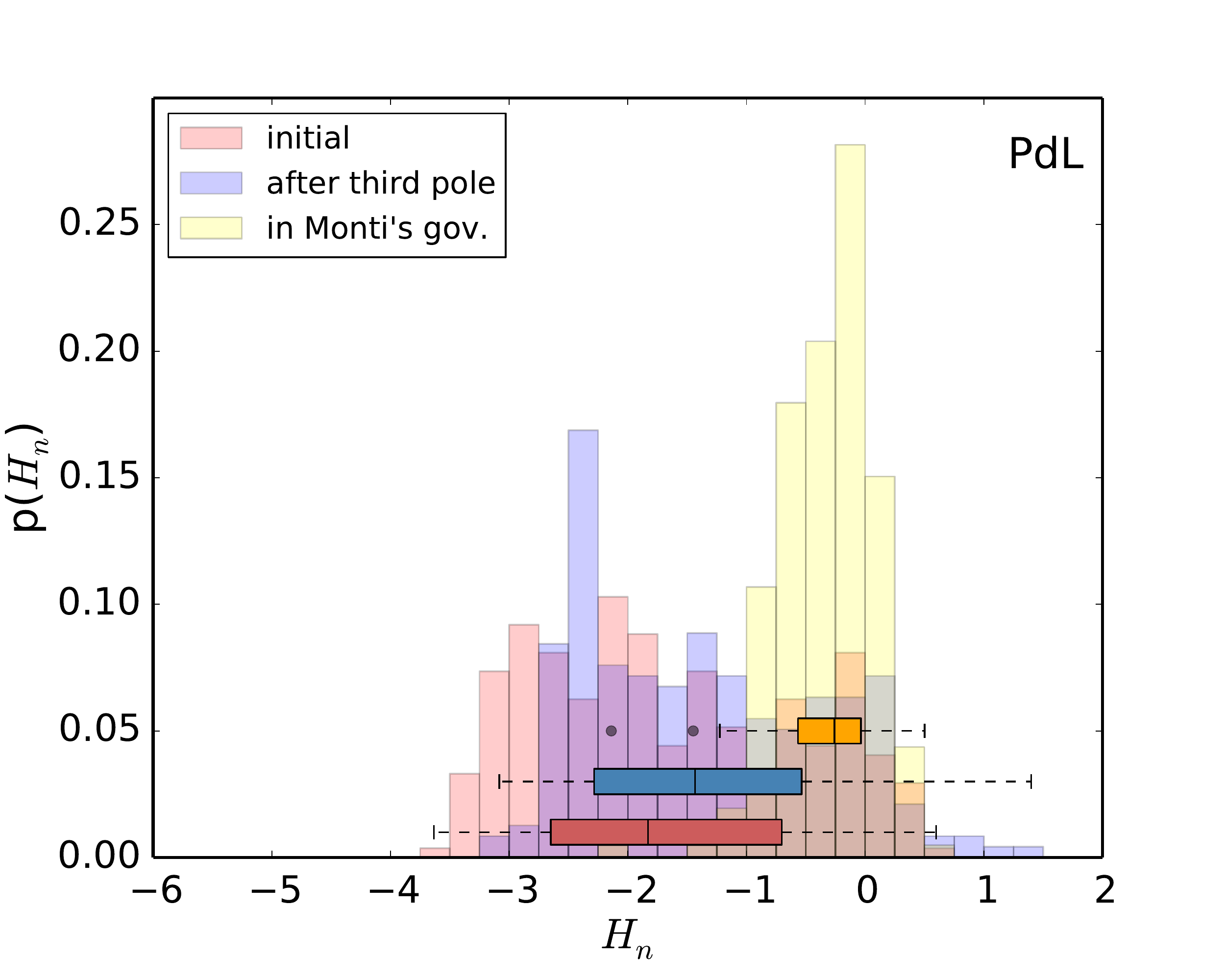} &
\includegraphics[width=0.5\textwidth]{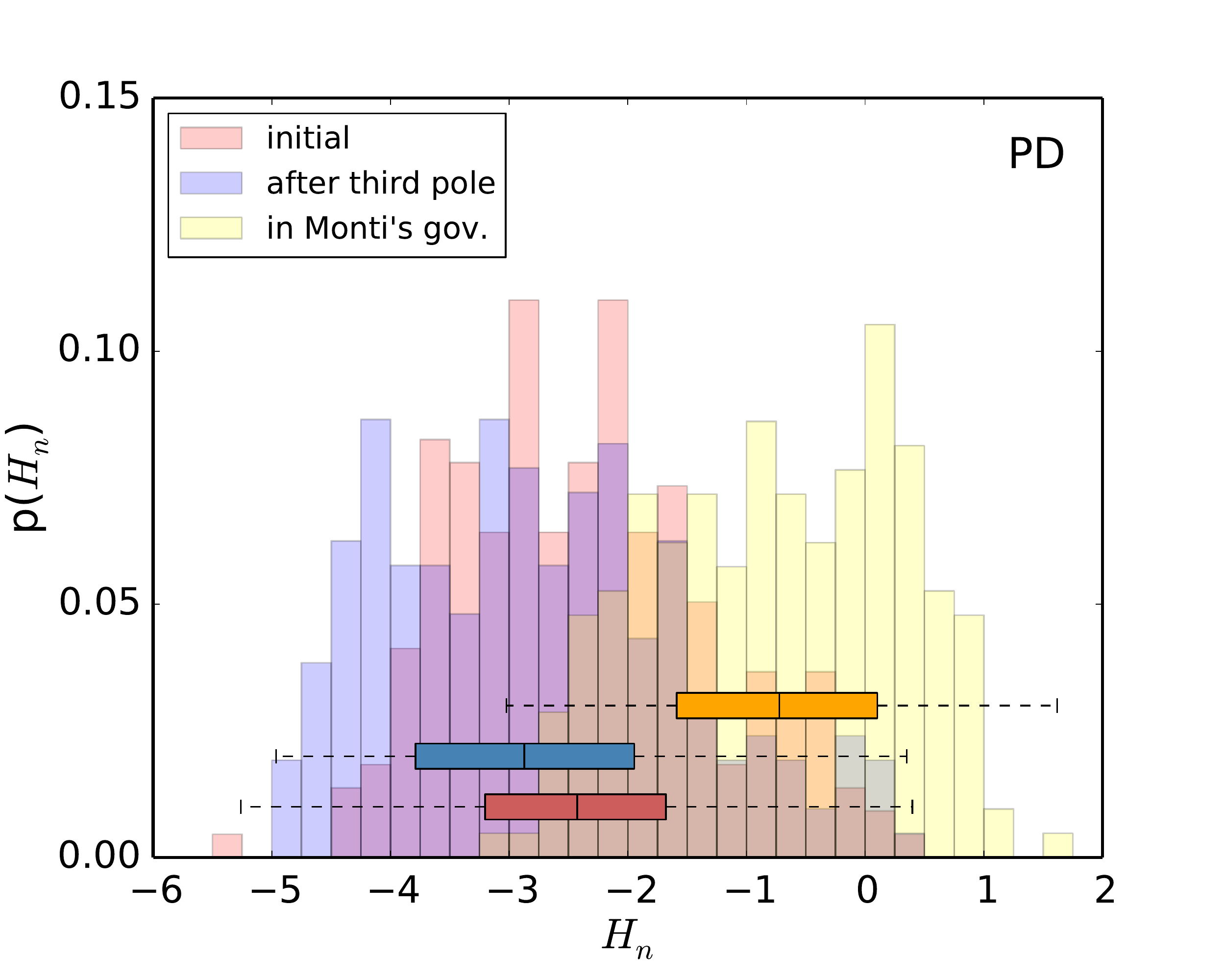} \\
(a) & (b)
\end{tabular}
\end{center}
\caption{(a) Distribution of the entropy for PDL MPs, with boxplots for the three periods: before Third Pole, from Third Pole to Monti's government establishment, during Monti's government. (b) Distribution of the entropy for each PD deputy, with boxplots for the same three periods.}
\label{fig:e3}
\end{figure}

\begin{figure}[!t]
\begin{center}
\includegraphics[width=\textwidth]{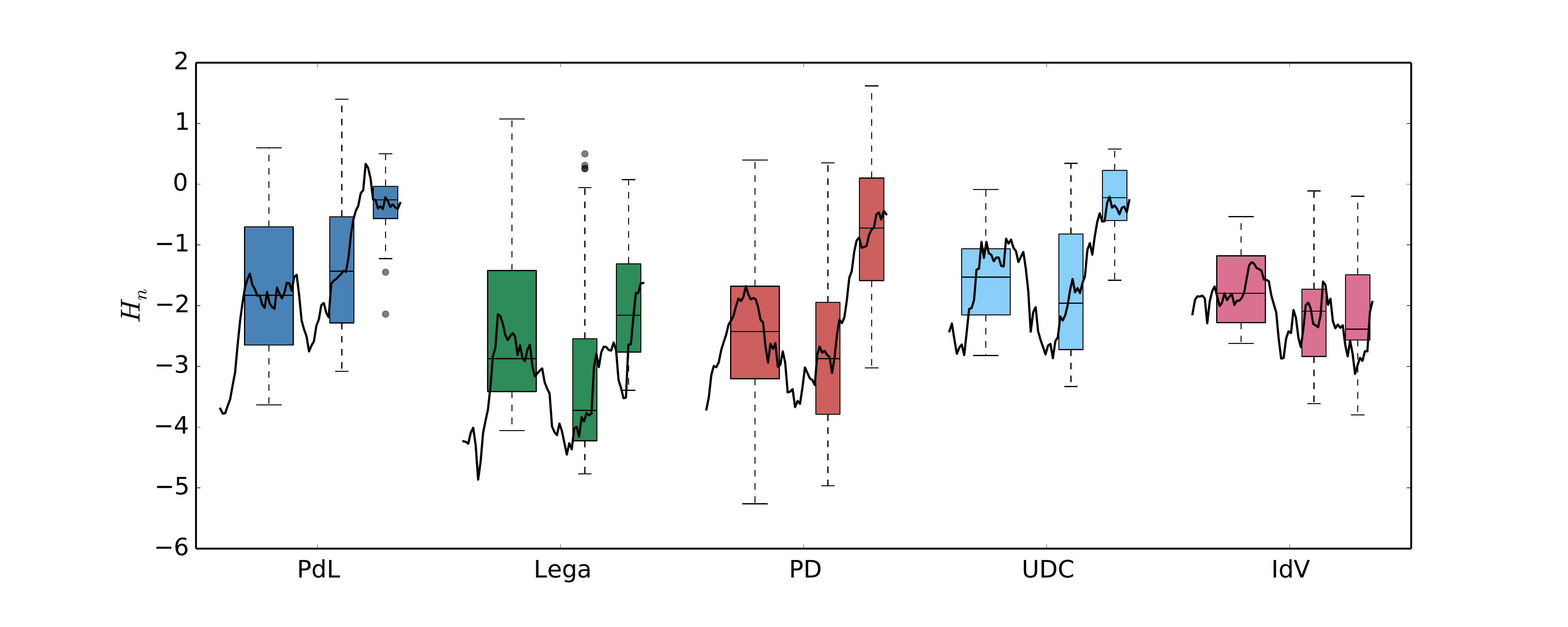}
\end{center}
\caption{Boxplots of the node entropy for the five main parties. Black lines show the median of the node entropy for each MP, over a sliding window of 10 timepoints.}
\label{fig:e4}
\end{figure}

To further support the claim of having three different time phases (in terms of voting behaviour) emerging from the group entropy analysis, we now focus on the node entropy by zooming on the behaviour of the single deputies.
In Fig.~\ref{fig:e3} we show the histograms representing the distribution of the node entropy for the MPs of PD and PDL, separately (this includes MPs later joining FLI only until point 44) and with different colors marking the three different phases with the corresponding boxplots.
In Fig.~\ref{fig:e4} complete trends and phase boxplots are shown for all the five major parties, with lines showing the median of the node entropy over a sliding window of 10 timepoints.
This analysis suggests the existence of the three phases and, further, the existence (for PD, Lega and UDC) of a "sling" effect, where the entropy decreases in the second phase before jumping to a close-to-zero value in the last time range. 
In Tab.~\ref{tab:stats} we list all the corresponding medians and the statistical significance of the mutual difference of the distributions (within each party for the three time phases) as $p$-value of one-tailed Kolmogorov-Smirnov 2-sample test.
Note that conversely in the PDL  case the 2-sample test shows an opposite direction with $p$-value $1.5\dot 10^{-5}$ between phase one and two, suggesting an increase in entropy after the internal division.
Interestingly, the sling effect is also not present for IdV, a party whose components slowly diluted into other groups during the legislature.

\begin{table}[!t]
\caption{(left) Median of the node entropy, stratified by parties and time ranges (P1: points 1-43, P2: points 44-61, P3:62-83); (right) $p$-values of the one-tailed Kolmogorov-Smirnov 2-sample test for the node entropy within each party compared to the null hypotesis that $Pj$ is not smaller than $Pi$ (Pj-Pi); $p$-value smaller than double precision are reported as zero.}
\begin{tabular}{ccc}
\begin{tabular}{l|rrr}
        & \textbf{P1} & \textbf{P2} & \textbf{P3} \\
\hline
\textbf{PDL}  &-1.83 &-1.43 &-0.26 \\
\textbf{Lega} & -2.87 &-3.72 &-2.16 \\
\textbf{PD} &  -2.43 &-2.87 &-0.72\\
\textbf{UDC} & -1.53 &-1.96 &-0.22\\
\textbf{IDV} & -1.80 &-2.09 &-2.39
\end{tabular}
$\qquad\qquad\qquad$
\begin{tabular}{l|rrr}
        & \textbf{P1-P2} & \textbf{P3-P1} & \textbf{P3-P2} \\
\hline
\textbf{PDL} & 1 & 0 & 0 \\
\textbf{Lega} & $6\cdot 10^{-5}$ & 0.0015 &  $4\cdot 10^{-10}$ \\
\textbf{PD} &$6\cdot 10^{-4}$ & 0 & 0 \\
\textbf{UDC} & 0.015 & $3\cdot 10^{-10}$ & $6\cdot 10^{-8}$\\
\textbf{IDV} & 0.079 & 0.875 & 0.508 
\end{tabular}
\end{tabular}
\label{tab:stats}
\end{table}

\section{Conclusions}
\label{sec:conclusions}
We introduced group and node alignment entropy as tool set for detecting partisan cohesion and individual coherence in political networks.
At the single snapshot level, our group entropy alignment measure can detect critical events associated with peaks of order mostly associated to forced party discipline. 
Over longer time periods group entropy may help revealing structural variations of the voting assembly.
The possibility of summarizing mesoscale dynamics with a single indicator suggests the potential use of the entropy as an input for change point detection algorithm for temporal segmentation.
Furthermore our definition of node alignment entropy, that measures how much a node swings with respect to the core of its community, contributes to the more general literature on node measures for temporal networks~\cite{cozzo2013clustering,kivela2013multilayer,halu2013multiplex} extending the scope of the paper beyond the specific case of political networks.

\bibliographystyle{unsrt}
\bibliography{lami14entropy}
\end{document}